\documentclass[12pt,preprint]{aastex}
\shorttitle{On the Dynamical Stability of the Solar System}
\shortauthors{Konstantin Batygin \& Gregory Laughlin}

\begin{document}
 
\title{On the Dynamical Stability of the Solar System}
\author{Konstantin Batygin\altaffilmark{1} \& Gregory Laughlin\altaffilmark{1,2} }
\affil{\altaffilmark{1}Lick Observatory, University of California, Santa Cruz, CA 95064}
\affil{\altaffilmark{2}Department of Astronomy and Astrophysics, University of California, Santa Cruz, CA 95064}
    
\begin{abstract}
A long-term numerical integration of the classical Newtonian approximation to the planetary orbital motions of the full Solar System (sun + 8 planets), spanning 20 Gyr, was performed. The results showed no severe instability arising over this time interval. Subsequently, utilizing a bifurcation method described by Jacques Laskar, two numerical experiments were performed with the goal of determining dynamically allowed  evolutions for the Solar System in which the planetary orbits become unstable. The experiments yielded one evolution in which Mercury falls onto the Sun at $\sim$1.261Gyr from now, and another in which Mercury and Venus collide in $\sim$862Myr. In the latter solution, as a result of Mercury's unstable behavior, Mars was ejected from the Solar System at $\sim$822Myr. We have performed a number of numerical tests that confirm these results, and indicate that they are not numerical artifacts. Using synthetic secular perturbation theory, we find that Mercury is destabilized via an entrance into a linear secular resonance with Jupiter in which their corresponding eigenfrequencies experience extended periods of commensurability. The effects of general relativity on the dynamical stability are discussed. An application of the bifurcation method to the outer Solar System (Jupiter, Saturn, Uranus, and Neptune) showed no sign of instability during the course of 24Gyr of integrations, in keeping with an expected Uranian dynamical lifetime of $10^{18}$ years.\end{abstract}


\section{Introduction}

One of the oldest, yet still relevant questions of astrophysics concerns the dynamical stability of the Solar System. This problem arose shortly after the introduction of Newton's Law of Universal Gravitation and has generated interest continuously since then. Classical perturbation theories developed by Lagrange and Laplace in the late 18th century (Laplace, 1799-1825), and later enhanced by Le Verrier and Newcomb (Le Verrier, 1856; Newcomb, 1891) provide a good approximation to the planetary motions over relatively short periods of time. Indeed, Laplace's success in explaining the secular motions of Jupiter and Saturn seemed to confirm the predictable clockwork of planetary orbits and helped to establish the long-standing philosophical concept of ``Laplacian determinism." In the late nineteenth century however, this world-view began to erode with Poincar\'e's demonstration that it is impossible to formulate an exact analytic solution for the motion of more than one planet, i.e. the non-integrability of the three-body problem (Poincar\'e, 1892). This advance foreshadowed the idea of chaotic systems and nonlinear dynamics.

Advances in computational technology and space travel have sparked continued interest in the problem of planetary orbits. In particular, it has been demonstrated that the Solar System displays chaotic behavior on sufficiently long time scales (Laskar 1989, 1990; Sussman \& Wisdom, 1992). The four terrestrial planets display chaotic motion with a Lyapunov time on the order of $\sim$5 million years, as do the Jovian planets (Laskar, 1989; Murray \& Holman, 1999 and the references therein). Furthermore, the mass ratios of the planets to the Sun are much larger than those required by the KAM theory to assure strictly bounded, quasi-periodic variations of the orbital elements (Arnold, 1961). It therefore appears that the Solar System may ultimately be dynamically unstable. If one waits long enough (ignoring drastic overall changes such as those wrought by the Sun's evolution or close encounters with passing stars), the planets may eventually find themselves on crossing orbits, which may lead to close encounters, ejections, or collisions. Our aim here is to investigate this possibility and estimate a characteristic time-scale.

\section{The Classical Problem of Solar System Evolution}

The Solar System can be modeled as a non-linear Hamiltonian N-body system, governed by Newton's law of universal gravitation:
\begin{equation}
\frac{d^2\textbf{\textit{r}}_i}{dt^2}=-G\sum_{i=1,j\not=i}^{N}\frac{m_{j}(\textbf{\textit{r}}_i-\textbf{\textit{r}}_j)}{|\textbf{\textit{r}}_i-\textbf{\textit{r}}_j|^3},
\end{equation}
where $\textbf{\textit{r}}$'s denote the positions of the bodies $i$ and $j$, $G$ is the gravitational constant, and $m$'s are the masses.
The gravitational potential does not exhibit an explicit time-dependence, so the Hamiltonian, $H$, corresponds to the sum of all kinetic and potential energies:
\begin{equation}
H=\sum_{i=1}^{N}\frac{\textbf{\textit{p}}_i^2}{2m_i} -G\sum_{i=1}^{N}m_{i}\sum_{j=i+1}^{N}\frac{m_{j}}{|\textbf{\textit{r}}_i-\textbf{\textit{r}}_j|},
\end{equation}
where $\textbf{\textit{p}}$ is the momentum.
Hamilton's equations of motion dictate the rates of change of the position and momentum of each body. 
\begin{equation}
\frac{d\textbf{\textit{r}}_{i}}{dt}=\frac{\partial{H}}{\partial{\textbf{\textit{p}}_{i}}}
\end{equation}
\begin{equation}
\frac{d\textbf{\textit{p}}_{i}}{dt}=-\frac{\partial{H}}{\partial{\textbf{\textit{r}}_{i}}}
\end{equation}
In our work here, all bodies are treated as point-masses. The Earth-Moon system is treated as a point-mass, located at the Earth-Moon barycenter.  The classical formulation ignores a host of high-order complications including general relativity, tidal friction, solar and planetary oblateness, solar mass loss, galactic tides, and the perturbative effect of passing stars. One could argue that by neglecting these effects, the resulting evolutionary trajectories are of little practical value. We feel however, that because of the chaotic nature of the system, it is worthwhile to make a probabilistic investigation of the stability of the classically defined Solar System. Historically, the general relativistic corrections have generated most attention, given their observability in the form of Mercury's excess precession of 0.43$^{\prime \prime}$/yr. We will consider the effects due to general relativity on the dynamical stability in section 5.

\subsection{Numerical Methods}

Mixed-variable symplectic integrators, such as the Wisdom-Holman map (Wisdom \& Holman, 1991), exhibit little long-term accumulation of energy error, beyond that due to round-off, and are considerably faster than other N-body algorithms (Chambers, 1999). This makes them a natural choice for long-term integrations. While the Wisdom-Holman symplectic mapping has proved to be both efficient and accurate, it breaks down when encounters between bodies take place (Wisdom \& Holman, 1991). We are interested in evolving the Solar System into configurations where orbit crossing occurs, so the employed algorithm must permit close encounters. The hybrid symplectic integrator, $\mathit{Mercury6}$, developed by Chambers (1999) is well situated for our analysis. This particular integration method is symplectic for all bodies, except those undergoing a close encounter.Close encounters are resolved via direct Bulirsch-Stoer integration, with the transfer occurring for encounters closer than 3 Hill radii (Chambers, 1999). In addition, we will also present a number of confirmation simulations, that used a straightforward Bulirsch-Stoer integration method (see Press et. al., 1992).

\section{Chaotic Phenomena and Lyapunov Exponents}

The chaotic motion of the planetary orbits is well established (Laskar, 1989). In keeping with the usual behavior of chaotic systems, small variations in planetary initial conditions produce large variations in the long-term evolution of the system. The degree of chaos can be measured by noting that if initially, two orbits differ by some small separation $s_{0}$ in phase space, this separation will grow exponentially with time. This relation amounts to the statement:
\begin{equation}
\frac{s}{s_{0}}= e^{(\gamma\Delta t)} ,
\end{equation}
where $\gamma$ is the Lyapunov characteristic exponent (Murray \& Dermott, 1999). The relation however, is only true for local divergence: if s becomes too large, equation (5) no longer provides a good model for the divergence of solutions and the degree of chaos intrinsic to a particular orbit. The Lyapunov exponent, $\gamma$, can easily be approximated numerically as an average of $\mathit{N}$ re-normalizations of the separation vector, at fixed time intervals $\mathit{\Delta t}$:
\begin{equation}
\gamma=\lim_{N\to\infty}\sum_{k=1}^{N}\frac{\ln{(\frac{s_{k}}{s_{o}})}}{N \Delta t}.
\end{equation}
The Lyapunov time, $1/\gamma$ is the duration required for $s_{0}$ to increase by a factor of \textit{e}. Table 1 presents the Lyapunov exponents and times, which we have obtained for all planets. The measurements were made with $\mathit{N}$ = 100, $\mathit{\Delta t}$ = 10,000 years. The initial displacements were all 150m, in the radially outward direction. Although it is hard to measure $\gamma$ within an accuracy of a factor of 2 (Murray \& Holman, 1999), our estimates of $\gamma$'s are in good quantitative agreement with values obtained by Laskar (1989) and Murray \& Holman (1999). Lyapunov times of order $\sim$5Myr signify that our ability to track the the planet's exact locations in phase space is lost on a time-scale far shorter than the age of the Solar System. Nevertheless, long-term integrations are still very useful in providing probabilistic evaluations of the Solar System's future behavior.

\section{Direct Long-term Integrations}

The fate of the Solar System can be sampled by numerically integrating the equations of motion over a very long time interval. This strategy, however, is known to be unlikely to produce a solution in which the planetary orbits will differ dramatically from their initial conditions. Many long-term solutions have been obtained via direct integrations (Tremaine, Duncan \& Quinn, 1991; Ito \& Tanikawa, 2002) as well as by integrations of averaged equations of motion (Laskar, 1989, 1990). Similarly, we  integrated the orbital planetary motions of the full Solar System for 20Gyr, starting from the current (DE 102, Newhall et al., 1983) configuration. Throughout the computation, we used a time-step of 8 days. The total energy of the system was conserved to $\Delta E/E$ $<$ 10$^{-7}$ and the total angular momentum was conserved to $\Delta L/L$ $<$ 10$^{-9}$. The planetary orbits showed only bounded, low-level excursion of their orbital elements. In particular, in concordance with secular theories, the semi-major axes stayed approximately constant throughout the integration. Nevertheless, we note that the long-term variations in Mercury's eccentricity exceed those of other planets. Figures (1) and (2) show the eccentricity evolutions of Earth and Mercury over 2 $\times$ 10$^{10}$ years. We reiterate that these results must be interpreted in light of the Solar System's chaotic nature. A long-term integration does not represent the actual behavior of the planets' motions over the integration time-interval, but rather a $\mathit{possible}$ trajectory, drawn from an enormous ensemble of outcomes (Laskar, 1994). Hence, as a demonstration that the Solar System is stable over its lifetime, this result is not satisfactory. Rather, it hints that any timescale on which instability might occur is likely to be very long.

\section{The Laskar Experiment}


A result published by Laskar (1994) presents a remarkable contrast with the apparent stability exhibited in our 20Gyr integration. Laskar noted that when a few-body system is to be evolved over many Lyapunov times, there is no need to compute a single continuous integration. Individual evolutionary paths, which differ through trivial changes in initial conditions, are all equally valid statistical representatives of the actual system.

Laskar performed the following experiment: using a fast numerical code which incorporated some $\sim$50,000 secular perturbation terms between the eight planets, he integrated the Solar System starting from today's configuration into negative time for 2 Gyr. He then repositioned Earth by 150m in four different directions, and integrated the four nearly identical variations of the Solar System further backwards in time for 500 million years (Laskar, 1996). Due to the highly chaotic nature of the system, for most of the computational time, each of Laskar's four simulations was exploring an entirely different dynamical path within the Solar System's allowed phase space. The behavior is analogous to the familiar ``butterfly's wings'' phenomenon in meteorology (Lorenz, 1963), and is wholly independent of the numerical method employed by the integrator. 

Laskar then examined the individual orbital histories and selected the trajectrory in which Mercury's eccentricity achieved its largest value. The Solar System configuration at the time of this greatest eccentricity excursion was then used as a starting condition for a second set of four individual 500 million year integrations. At the end of this second round of calculations, a new set of starting conditions was determined by again selecting the configuration at which Mercury's excursion was the largest (Laskar, 1994).  

After 18 rounds, which when pieced together yielded a 6Gyr integration, Mercury's eccentricity increased above $e>0.5$. The high eccentricity led to much stronger variations in orbital elements, delivering Mercury into a zone of increased chaos, where it was in danger of suffering a close encounter with Venus, a collision with the Sun, or an ejection from the Solar System. Laskar also reported a second experiment done in positive time, and using the same method. In the second experiment, perturbations in Earth's position were only 15m and it  took 13 rounds, involving a trajectory integration time of 3.5Gyr to destabilize Mercury (Laskar, 1994). Henceforth, we will refer to this general approach as the ``Laskar method," and any experiment utilizing this method as a ``Laskar experiment."

Laskar's composite trajectory, which leads directly to instability appears to be the first explicit demonstration of the Solar System's long-term dynamical instability. In light of this unexpected result, there are a number of very interesting unanswered questions. First, was the escape of Mercury a consequence of the secular perturbation approach which Laskar employed in order to rapidly complete 500 million year integrations? Would his bifurcation strategy find a similar result when used with direct numerical integration of the equations of motion? 

If so, what is the dynamical mechanism that destabilizes the inner Solar System? One can imagine a particular integration of the Solar System as a state vector sweeping through a large parameter space, the allowed regions of which are constrained by the conservation of the energy and angular momentum of the entire system.  A fraction of the overall parameter space consists of states in which the planets are on crossing orbits. With its sensitivity to evolving orbital parameters, which route will the bifurcation strategy take to reach a state where orbit crossings occur?

Finally, if the Laskar method is indeed a computationally efficient way of finding trajectories to unstable configurations of the Solar System, what is the extent of its general utility? That is, can it harvest chaotic diffusion to discover possible evolutions of the system which lead \textit{any given planet} into a zone of instability, given a large number of Lyapunov times? These are the questions which we are proposing to answer. 

We have studied the effect of replacing computationally efficient (but still approximate) secular theory with direct numerical integration on Laskar's experiments. First, a direct integration spanning 500Myr ($\sim$100 Earth Lyapunov times), with unchanged initial conditions was performed. Picking up at this endpoint, five solutions for 500Myr were computed, four of these had Earth's position shifted, and one with Earth's position unchanged. The four perturbations were provided in the ecliptic plane. Because initial uncertainties diverge exponentially with time, a shift of 150m in Earth's position 500Myr into the future, corresponds to an initial error of order 10$^{-42}$m i.e. 10 orders of magnitude smaller than the Plank scale, and far smaller than our numerical resolution. The solution in which Mercury attained the highest eccentricity was preserved to the nearest whole Myr, and the five bifurcations were started again. Figure (3) shows a generalized flow-chart of the strategy. 

In keeping with Laskar's example, two experiments were performed: one with Earth's position shifted by 150m during the bifurcation process and the other, by 15m. Both experiments were done in positive time. In the following, we shall analyze the results of the 150m experiment in deeper detail because Mercury's transition from stable to unstable motion can be seen in a more pronounced manner than that in the 15m experiment.

\subsection{Time-step and Numerically-induced Chaos}
When obtaining numerical solutions for a chaotic system, it is often easy to confuse real dynamical instabilities with numerically generated instabilities. The Solar System is a Hamiltonian system, in which the total energy and total momentum are conserved. When an integrator fails to respect these conservations, numerically-induced chaos arises and the solution becomes unreliable (Yoshida, 1993). In the particular case of the Solar System, the limiting factor is the requirement of accurately resolving Mercury's orbit. An 8-day time-step (Ito \& Tanikawa, 2002) has been shown sufficient for integration of the Solar System as long as Mercury's eccentricity does not rise significantly above its $e\sim0.2$ current value. Since we are anticipating the possibility of Mercury's eccentricity to attain values near unity, the time-step will have to be reduced as Mercury's eccentricity grows. 

We enforced strict requirements concerning conservation of the integrals of motion on the integration process. We have observed through trial and error that in order for numerical error to cause significant non-physical variations in Mercury's orbital elements, the energy or angular momentum non-conservation must fall above $\Delta E/E$ $\sim$ 10$^{-5}$ or $\Delta L/L$ $\sim$ 10$^{-7}$, respectively. Thus, the particular values we chose require 3 orders of magnitude better conservation than the ``critical" values, while still allowing for realistic integration time-steps. Specifically, at any given point in time, total energy of the system must be conserved to an order of one part in one hundred million ($\Delta E/E$ $\lesssim $ 10$^{-8}$), while total angular momentum must be conserved to one part in ten billion ($\Delta L/L$ $\lesssim $ 10$^{-10}$). If on any occasion, this requirement was violated, the particular step of the Laskar method was recomputed with a reduced time-step as to satisfy this requirement. The time-step used in the 150m and the 15m Laskar experiments was varied from 3 days to 1.2 days. Figures (4) and (5) show the fractional change in the Solar System's total energy and angular momentum respectively, for the 15m Laskar experiment. Figures (6) and (7) show the same for the 150m Laskar experiment. The dashed lines indicate the times when the bifurcations were applied. Subsequently, they also indicate reductions of time-step of the integrator. The last Laskar step in both experiments was computed entirely using the Bulirsch-Stoer algorithm, thus the change in conservation properties. 

\subsection{Results and Analysis}
We observed the loss of Mercury from the Solar System in both our 15m and the 150m Laskar experiments. Tables 2 and 3 present the step-by-step progressions for these two experiments. In the 15m Laskar experiment, Mercury was destabilized after four steps. Its eccentricity gradually rose to approximately 0.45 before suffering a large increase at $t\sim$ 1Gyr. Mercury then collided with the Sun at 1.261Gyr. That is, due to a very high eccentricity, Mercury's came within $r_{min} < 0.005$ AU of the Sun's center. Figure (8) shows the complete evolution of Mecury's eccentricity as a function of time for this experiment. Again, dashed lines indicate the times when the Laskar method was applied. Figures (4) and (5) show the fractional conservation of total energy and angular momentum for this experiment. In the 150m experiment, Mercury was destabilized even faster: only three steps were needed to stimulate Mercury to collide with Venus at $t\sim$ 862Myr. Figure (9) presents Mercury's eccentricity as a function of time for this experiment. Figure (10) presents the minimal distance of approach during a final series of close encounters between Mercury and Venus. The collision takes place at $t \sim$ 861.455Myr, when $d_{min} = 5.5561\times10^{-5}$ AU $< r_{venus} +r_{mercury} = 5.6762 \times 10^{-5}$ AU. In addition to Mercury, Mars was also ejected from the Solar System in the 150m Laskar experiment. This happened at $t\sim$ 822Myr, shortly after Mercury's entrance into a zone of severe chaos. This was facilitated by a sudden spike in Mars's eccentricity, which can be observed at the end of its orbital evolution in Figure (11), as well as a similar spike in Mars' semi-major axis. Mars' ejection from the Solar System was assumed after the magnitude of the radial vector exceeded $r_{max} > 100$ AU. The fractional energy and angular momentum conservations for the 150m Laskar experiment are presented in Figures (6) and (7). Note that for both of these numerical experiments, the conservations of the integrals of motion strictly follow the requirements implemented in section 5.1.

The behavior of Mercury's orbital elements for the two solutions is identical for the first 500Myr because the first steps of the composite integrations were started from the same (current) initial conditions. Nevertheless, Mercury's orbital evolution which lead up to Mercury's collision with the sun (as in the 15m experiment), or its collision with Venus (as in the 150m experiment) was entirely different for the two experiments. We can thus conclude that a wide array of unstable solutions exist for the Solar System's future. While there is a certain vicarious thrill in tracking Mercury's highly unstable mode of behavior in detail, an area of deeper importance lies in understanding of why and how readily Mercury makes a transition from regular to irregular motion. To investigate the nature of this process, we conducted multiple tests to evaluate the validity of the results.

An immediate question is whether the development of instability is associated with either the time-step or the integration algorithm. We accomplished these tests by reintegrating a 22Myr time interval of the system's evolution found in the 150m experiment, during which transition from stable to unstable motion takes place. The principal solution is plotted as the solid blue curve over this time interval in Figure (12). We first integrated with a time-step of 0.5 days using the symplectic algorithm and then did the same with the Builch-Stoer algorithm, also setting its time-step to 0.5 days and its convergence tolerance to 1 $\times$ 10$^{-15}$. Both of these integrations yielded almost identical increases in eccentricity as our primary solution. The same test was performed with the results of the 15m experiment. The reduced time-step symplectic and Bulirsch-Stoer replication integrations were started from 1190Myr, prior to the appearance of any unusual orbital eccentricities, and spanned 20Myr. As with the 150m experiments, the results were practically indistinguishable from the principle solution. In both cases, Mercury made the transition to the dynamical regime where it eventually became unstable. Consequently, we have significantly increased confidence that this result is indeed physical and is not a numerical artifact.

Next, we examined if these result can be altered by providing small perturbations to Mercury, prior to its increase in eccentricity. If Mercury's motion is in the vicinity of a separatrix, tiny perturbations will result in entirely different evolutions (Thornton \& Marion, 2004). Working with the results of the 150m experiment, starting at 778Myr, we provided perturbations of 15m and 150m magnitude to Mercury in each of the four directions and integrated eight such perturbed solutions for 22Myr, spanning the transition to unstable motion. Again however, these integrations reproduced our primary solution very closely, showing that the increase in Mercury's eccentricity was already implicit in the system's configuration at 778Myr. The exact same procedure was followed in conducting this test on the 15m experiment: Mercury was perturbed by 15m and 150m in four different directions at 1190Myr and the eight variations of the solution were  integrated for 20Myr. Once more, the original solution proved to be quite deterministic in its rise in eccentricity.

One can further ask: what is causing Mercury's eccentricity to escalate? Let us consider the 150m experiment, since the transition from stable motion to unstable motion is more clearly expressed in this example. If we examine the progression of Mercury's eccentricity (Figure 9), it is clear that up until 780Myr, its eccentricity varies within a narrow and well defined range, suggesting that the evolution is well described by quasi-periodic motion. However shortly thereafter, Mercury's eccentricity begins to suffer an almost linear increase (Figure 12) . This increase in eccentricity leads to Mercury's eventual dramatic misadventure.

The chaotic motion of the outer Solar System has been shown to arise from overlapping mean motion resonances among the four Jovian planets (Murray \& Holman, 1999). Mercury however, does not appear to participate in any mean-motion resonance throughout the transition from stable to unstable motion. That is, we were unable to identify any librating arguments associated with mean-motion resonances. Instead, secular resonance serves as a driver for the sudden increase in Mercury's eccentricity. 

\subsubsubsection{Synthetic Secular Theory}

A hint of Mercury's participation in a secular resonance can be seen in a comparison of Mercury's disturbing function which corresponds to stable motion, and that which corresponds to unstable motion. The second order averaged secular disturbing function for planet $i$ can be written as
\begin{equation}
\left< \mathcal{R}_{i}^{sec} \right> =  n_{i}a_{i}^{2}\left[\frac{1}{2} A_{ii} e_{i}^{2} + \frac{1}{2} B_{ii} I_{i}^{2} +\sum_{j=1,j\neq{i}}^{N}\left(A_{ij}e_{i}e_{j}\cos(\varpi_{i}-\varpi_{j})+B_{ij}I_{i}I_{j}\cos(\Omega_{i}-\Omega_{j})\right)\right], 
\end{equation}
where $n$ is the mean motion, $a$ is the semi-major axis, $e$ is the eccentricity, $I$ is the inclination, $\varpi$ is the longitude of pericentre, $\Omega$ is the ascending node, and the $A$'s and $B$'s are constants which depend on masses and semi-major axes only (Murray \& Dermott, 1999). If we adopt a synthetic secular theory (see Malhotra et al, 1989; Laskar, 1990), the dominant frequencies of the secular disturbing function can be identified numerically by Fourier analyzing the numerically computed (equation 8) time-series for Mercury's full disturbing function. By definition, secular terms are slowly varying and will have low frequencies. To the extent that Mercury's motion is controlled by the approximation inherent in equation (7), the numerically measured disturbing function,
\begin{equation}
\mathcal{R}_{i}=\sum_{j=1,j\neq{i}}^{N} \left(\frac{Gm_{j}}{|\textbf{\textit{r}}_j-\textbf{\textit{r}}_i|}-Gm_j\frac{\textbf{\textit{r}}_{i}\cdot\textbf{\textit{r}}_{j}}{|\textbf{\textit{r}}_j|^3}\right)
\end{equation}
will agree with the above expression.

Having numerically obtained the time-series for the longitudes of pericentre, $\varpi_{i}$, of all the orbits, the secular eigenfrequencies, $g_{i}$, of the system, and the frequencies of the secular disturbing function, ($g_{i} - g_{j}$) can identified. Physically the eigenfrequencies of the system correspond to averaged orbital precession rates,  $g_{i} =  \left< \dot{\varpi}_{i} \right> $. An examination of the time-series shows that the particular frequency of interest for us is ($g_{1} - g_{5}$). The dashed line in Figure (13) denotes its location in the frequency spectrum of Mercury's disturbing function which corresponds to stable motion. For a stable configuration, this frequency is $(g_{1} - g_{5}) \approx 0.9389 ^{\prime \prime}$/yr. When we look for ($g_{1} - g_{5}$) in the unstable disturbing function, (Figure 14) we quickly note that this frequency has shifted to $(g_{1} - g_{5}) \approx 0.0538 ^{\prime \prime}$/yr, much closer to zero. Essentially, this signals that Mercury's and Jupiter's orbits evolved to have nearly equal averaged precession rates.

\subsubsubsection{Secular Resonance}
Treating Mercury as a planet of negligible mass, the classical Laplace-Lagrange secular solution for Mercury's eccentricity vector ($h_{1}=e_{1}\sin\varpi_{1}$, $k_{1}=e_{1}\cos\varpi_{1}$) is given by
\begin{equation}
h_{1}=e_{1free}\sin(g_{1}t+\beta_{1})-\sum_{j=2}^{8}\frac{\nu_{j}}{g_{1}-g_{j}}\sin(g_{j}t+\beta_{j}),
\end{equation}   
\begin{equation}
k_{1}=e_{1free}\cos(g_{1}t+\beta_{1})-\sum_{j=2}^{8}\frac{\nu_{j}}{g_{1}-g_{j}}\cos(g_{j}t+\beta_{j}).
\end{equation}   
In equations (9) and (10), $\nu_{j}$'s are the components of the eigenvectors of the \textbf{A} matrix, multiplied by constants which depend only on masses and semi-major axies. The quantity $e_{1free}$ and the phase constants, $\beta_{1}$ and $\beta_{j}$'s, are determined by initial conditions (Murray \& Dermott, 1999). The frequency $g_{1}$ is Mercury's proper frequency, while the rest are its forcing frequencies. 

A resonance arises when two or more oscillators have frequencies which are in simple numerical ratio. A linear secular resonance occurs when a planet's proper frequency is approximately  equal to one of the forcing frequencies. If this occurs, the denomenators in equations (9) and (10) become small, causing large variations in a particle's eccentricity vector (Namouni \& Murray, 1999). In classical perturbation theory, all $g_{i}$'s are fixed in time, however in reality, they vary (Laskar, 1990). As stated before, these frequencies and their variations can be estimated numerically. In the classical approximation used here, the current value of Mercury's proper frequency is $g_{1current} = 5.4058 ^{\prime \prime}$/yr (Newcomb, 1891). During Mercury's rapid climb in eccentricity, its proper frequency evolves to become ($g_{1} \approx 4.2973 ^{\prime \prime}$/yr) very close to Jupiter's forcing frequency ($g_{5} \approx 4.24354^{\prime \prime}$/yr). The evolution of ($g_{1}$ - $g_{5}$) resonance can be understood by looking at the evolution of the ($\varpi_{1}$ - $\varpi_{5}$) apsidal angle. If we examine ($\varpi_{1}$ - $\varpi_{5}$) as a function of time (Figure 15), we notice that until 787.5Myr, it circulates slowly from -2$\pi$ to +2$\pi$. Qualitatively, this reflects the two large-amplitude oscillations in Mercury's eccentricity, observed in Figure (12). However, ($\varpi_{1}$ - $\varpi_{5}$) librates between +19.8 and -43.56 degrees from approximately 787.5Myr to 790.5Myr. The apsidal angle's libration around a net-negative value forces a nearly linear increase in Mercury's eccentricity, causing it to rise to $e>0.6$. The absence of an analogous subsequent libration around a net-positive value results in the lack of recovery of Mercury's eccentricity. 

The secular behavior of Mercury's eccentricity due to the ($\varpi_{1}$ - $\varpi_{5}$) argument can be identified  explicitly through Lagrange's equations of planetary motion. The rate of change of eccentricity is given by
\begin{equation}
\frac{de}{dt}=-\frac{\sqrt{1-e^{2}}}{na^{2}e} (1-\sqrt{1-e^{2}})\frac{\partial{\mathcal{R}}}{\partial{\epsilon}} -\frac{\sqrt{1-e^{2}}}{na^{2}e}\frac{\partial{\mathcal{R}}}{\partial{\varpi}},
\end{equation}
and the rate of change of semi-major axis is given by
\begin{equation}
\frac{da}{dt}=\frac{2}{na}\frac{\partial{\mathcal{R}}}{\partial{\epsilon}},
\end{equation}
where $\mathcal{R}$ is the disturbing function, and $\epsilon$ is the mean longitude at epoch (Murray \& Dermott, 1999). However, during the time-interval of interest, to a very good approximation, Mercury's semi-major axis remains constant. Therefore $\partial{\mathcal{R}_{1}} / \partial{\epsilon} \propto da_{1}/dt \approx 0.$ Let us now consider Mercury's secular interaction with Jupiter. Extracting only the ($\varpi_{1}$ - $\varpi_{5}$) resonant term from the disturbing function, the resulting rate of change in Mercury's eccentricity is given by:
\begin{equation}
\frac{de_{15}^{*}}{dt} \approx A_{15} e_{5} \sqrt{1-e_{1}^{2}}\sin(\varpi_{1}-\varpi_{5}).
\end{equation}
Using the time-series of the eccentricities and longitudes of pericentre of Mercury and Jupiter, obtained from numerical integration, a time-series for $de_{15}^{*}/dt$ was constructed using equation (13). The $de_{15}^{*}/dt$ time-series was then numerically integrated to generate a time-series for $e_{15}^{*}$. It is important to keep in mind that $e_{15}^{*}$ results only from the ($\varpi_{1}$ - $\varpi_{5}$) argument in the disturbing function. Since $\dot{\varpi_{1}}$ is dominated by $g_{1}$, and $\dot{\varpi_{5}}$ is dominated by $g_{5}$ (Laskar, 1990), this solution is roughly equivalent to singling out the qualitative behavior of Mercury's eccentricity due to the ($g_{1} - g_{5})$ resonance. This solution is plotted as a dashed line in Figure (12), and resembles a low-amplitude, low-pass filtered version of the full solution.

The partial solution for $e_{15}^{*}$ shows explicitly that ($g_{1} - g_{5}$) secular resonance is responsible for the climb of Mercury's eccentricity, and ultimately the destabilization of the Solar System. However, as can be seen from Figure (12), when only the Jupiter-Mercury interaction is considered, the actual amplitude of variations in Mercury's eccentricity due to this argument are significantly smaller than that of the full solution. Indeed Mercury's eccentricity varies mostly due to interactions with Venus. Unlike $\dot{\varpi_{5}}$, $\dot{\varpi_{2}}$ is not dominated by $g_{2}$ (Laskar, 1990). It must be stressed that although a given eigenfrequency $g_{i}$ corresponds to the averaged precession rate of the $i^{th}$ planet, the secular solution for the longitude of pericentre
\begin{equation}
\varpi_{i}=\tan^{-1}\left( \frac {\sum_{j} e_{ij} \sin( g_{j} t+ \beta_{j})}{\sum_{j} e_{ij} \cos( g_{j} t+ \beta_{j})} \right)
\end{equation}
is a superposition of all eigenfrequencies. Figure (16) shows the Fourier analysis of $\varpi_{2}$. Indeed, there is a strongly pronounced presence of $g_{5}$ in $\dot{\varpi_{2}}$ and note that $g_{5}$'s amplitude is approximately equal to that of $g_{2}$. Following the same process as before, a partial solution for $e_{12}^{*}$ was constructed. The solution for $e_{12}^{*}$ is plotted as a solid black line in Figure (12). If $e_{12}^{*}$ and $e_{15}^{*}$ are added together, the full numerically obtained evolution of Mercury's eccentricity (blue curve in Figure 12) is recovered very closely. 

\subsubsection{Effects due to General Relativity}
The classical problem of the stability of the Solar System relies on the assumption of Newtonian gravity. Up to this point, we have been ignoring the effects of general relativity. Nevertheless, when we consider the dynamics of the inner Solar System, especially that of Mercury, the effects of general relativity play a significant role. Since the source of instability has been identified as a secular resonance in longitudes of pericentre, we can anticipate that the sudden addition of the effects of general relativity into the model will destroy the resonance, by adding 0.43$^{ \prime \prime }$/yr to Mercury's precession rate. The leading-order effects of general relativity can be modeled by adding an extra term to the sun's gravitational potential of the form
\begin{equation}
V _{GR}=-G \frac{M _{\odot} \ell^2}{c^2 r^3},
\end{equation}
where $\mathit{c}$ is the speed of light, $\mathit{r}$ is the radial distance of the planet, $\mathit{M_{\odot}}$ is the mass of the sun and $\ell$ is the orbital angular momentum per unit mass. We used the Bulirsch-Stoer method to integrate the equations of motion with the inclusion of the post-Newtonian potential term given above.  As expected, adding this term to the model completely eliminates all previously existing instabilities. Figure (17) shows the stable behavior of Mercury's eccentricity, computed as described, over the time-interval during which Mercury is destabilized, when the classical model is employed. This however does not imply that general relativity completely stabilizes the Solar System, since its effects were added to a system which has evolved to become unstable as a purely Newtonian object. Rather, we predict that it will prolong Mercury's dynamical lifetime. As stated before, a stable (current) configuration of the solar system, utilizing Newtonian gravity has $g_{1Newtonian} = 5.4058 ^{\prime \prime}$/yr and $g_{5} =  4.2435 ^{\prime \prime}$/yr. Account of general relativity moves $g_{1}$ to $5.8358 ^{\prime \prime}$/yr. Therefore, neglecting general relativity cuts down the distance between $g_{1}$ and $g_{5}$ by some 39\%. This is likely to allow Mercury to enter the ($g_{1}$ - $g_{5}$) secular resonance faster. 

Although our results are qualitatively similar to those of Laskar's, there is an interesting quantitative distinction. Laskar's experiment required 13 steps and 3.5Gyr to destabilize Mercury, whereas our experiment required 3 such steps, and less than one billion years for Mercury to collide with Venus. This could be due to our approach of utilizing full equations of motion, reflecting the chaotic nature of the Solar System more strongly than the perturbation theory approach, employed by Laskar. On the other hand, this could be due to the absence of general relativity in our model, while Laskar's model takes its effects into account (Laskar, 1996). In any event, a better understanding of the role that general relativity plays in the stability of the Solar System, can be attained by conducting another set of Laskar experiments using a model which accounts for its effects.

\subsection{Laskar Experiment for Uranus}
It appears that the Laskar method enhances our ability to sample the more chaotic of the dynamically allowed trajectories for a given planet. One then wonders if a highly chaotic solution can be found for any planet by utilizing the Laskar method. Murray \& Holman (1999) estimated that due to overlapping mean motion resonances involving Jupiter, Saturn and Uranus, the dynamical lifetime of Uranus is of order ~10$^{18}$ years. Hence, we decided to investigate whether an explicit trajectory leading to destabilization of Uranus can be found with the Laskar method. For this experiment, the model for the Solar System was simplified to include only the four Jovian planets, with the total mass of the terrestrial planets added to the Sun. The procedure was very much the same as before. This time however, Jupiter was perturbed instead of Earth. The amplitude of the perturbation was 1500m and each bifurcation step lasted for some 5Gyr. The progression of the Uranus Laskar experiment is presented in Table 4.

After ten Laskar steps, spanning 24Gyr, Uranus's eccentricity never exceeded $e = 0.078$. Figure (18) shows Uranus's eccentricity as a function of time.  Our inability to destabilize Uranus suggests that the Laskar method provides less than a $\sim10^{7}$ factor speed-up in the estimated dynamical lifetime. Clearly, an interesting step for future work is to provide a more meaningful calibration of the the method. Such calibration will be helpful in providing a real interpretation of the meaning of our full system experiments. 

\section{Discussion}

The numerical experiments and the accompanying analysis raise some very interesting issues. 

First, what has been accomplished? We have verified the general conclusions of Laskar's bifurcation experiments, and in so doing, have provided a dynamically consistent integration of Newton's equations of motion which both (1) start from initial conditions that are fully consistent with the current Solar System configuration, and (2) evolve the planets to a situation in which orbit crossing occurs. The confidence in this finding's validity is greatly enhanced by the recent determinations of Laskar (2008), who reached the same basic result independently.

What are the consequences of the obtained solutions? Our results underscore the realization that the inner and outer Solar System are prone to fundamentally different modes of instability. In the outer Solar System, it is the presence of overlapping mean-motion resonances that will eventually lead to instability. This however, will not take place within any reasonable time-frame. In the inner Solar System, disaster is brought on by the ($g_{1}-g_{5}$) secular resonance. The time-scale for this to occur is potentially less than the Sun's remaining lifetime, but the effects of general relativity will have to be considered in greater detail to validate this result.

Although our results provide an approximate lower limit on the dynamical lifetime of the inner Solar System of $T_{min} \sim 10^{9}$ years, a fundamental question remains unanswered: what is the \textit{expected} dynamical lifetime of the inner Solar System? Equivalently, what are the odds that the planets will evolve onto crossing orbits prior to the Sun's red giant phase? Laughlin \& Adams (2000) used a Monte-Carlo approach to evaluate the possibility that Earth will be ejected into interstellar space as a result of a chance encounter with a passing stellar system and found an overall probability of order $2 \times 10^{-5}$ that Earth will find its orbit seriously disrupted by this process within the next 5Gyr. Apparently, the mechanism studied in this paper adds significantly to this probability, as Laskar (2008) estimates a 1 - 2 \% chance that Mercury's eccentricity will rise above 0.6 in the same time-frame. In conclusion, it would seem that the Solar System is indeed not as stable as once thought. Subsequently, further studies of this issue, utilizing more precise physical models are greatly encouraged.

\begin{acknowledgments}{}
We thank John Chambers, Yuri Batygin, Matt Holman and Eugenio Rivera for helpful conversations and advice on computational methods. We would also like to thank the anonymous referee for insightful comments. This research was funded by NSF Career Grant \#AST-0449986 and NASA Planetary Geology and Geophysics Program Grant \#NNG04GK19G to Greg Laughlin.
\end{acknowledgments}{}

{}

\clearpage

\begin{table}
\begin{center}
\caption{Lyapunov exponents and times for the Solar System}
\begin{tabular}{lrrrr}
\tableline\tableline
Planet & Lyapunov exponent (years$^{-1}$) & Lyapunov time (years) &  \\
\tableline
Mercury &7.32029 $\times$ 10$^{-7}$ & 1.36607 $\times$ 10$^{6}$ \\
Venus &1.38561 $\times$ 10$^{-7}$ & 7.21703 $\times$ 10$^{6}$ \\
Earth &2.07484 $\times$ 10$^{-7}$ & 4.81964 $\times$ 10$^{6}$ \\
Mars &2.22353 $\times$ 10$^{-7}$ & 4.49736 $\times$ 10$^{6}$ \\
Jupiter &1.19528 $\times$ 10$^{-7}$ & 8.36623 $\times$ 10$^{6}$ \\
Saturn &1.56875 $\times$ 10$^{-7}$ & 6.37452 $\times$ 10$^{6}$ \\
Uranus &1.33793 $\times$ 10$^{-7}$ & 7.47423 $\times$ 10$^{6}$ \\
Neptune &1.49602 $\times$ 10$^{-7}$ & 6.68440 $\times$ 10$^{6}$ \\

\tableline
\end{tabular}
\end{center}
\end{table}

\begin{table}
\begin{center}
\caption{Progression of the 150m Laskar experiment}
\begin{tabular}{lll}
\tableline\tableline
Step number & Time interval (Myr) & End-point $e_{Mercury}$ \\
\tableline
1 & $0 - 500$     &  $0.2907$\\
2 & $500 - 797$ & $0.4391$\\
3 & $797 - 862$ & $0.8257$\\
\tableline
\end{tabular}
\end{center}
\end{table}

\begin{table}
\begin{center}
\caption{Progression of the 15m Laskar experiment}
\begin{tabular}{lll}
\tableline\tableline
Step number & Time interval (Myr) & End-point $e_{Mercury}$ \\
\tableline
1 & $0 - 500$     &  $0.2907$\\
2 & $500 - 994$ & $0.4139$\\
3 & $994 - 1207$ & $0.4874$\\
4 & $1207 - 1261$ & $ 0.9751$\\
\tableline
\end{tabular}
\end{center}
\end{table}

\begin{table}
\begin{center}
\caption{Progression of the Uranus Laskar experiment}
\begin{tabular}{lll}
\tableline\tableline
Step number & Time interval (Gyr) & End-point $e_{Uranus}$ \\
\tableline
1 & $0 - 5$     &  $0.0162$\\
2 & $5 - 5.78$ & $0.0365$\\
3 & $5.78 - 5.782$ & $0.0167$\\
4 & $5.782 - 8.673$ & $0.0551$\\
5 & $8.673 - 11.616$ & $0.0293$\\
6 & $11.616 - 13.694$ & $0.0780$\\
7 & $13.694 - 13.695$ & $0.0312$\\
8 & $13.694 - 14.345$ & $ 0.0318$\\
9 & $14.345 - 18.927$ & $ 0.0498$\\
10 & $18.927 - 24$ & $0.0605 $\\

\tableline
\end{tabular}
\end{center}
\end{table}

\clearpage

\begin{figure}
\centering
\includegraphics[width=450pt]{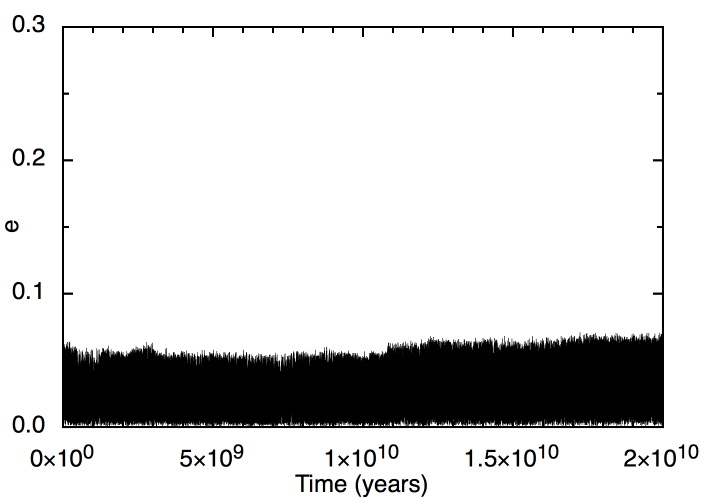}
\caption{Eccentricity of Earth as a function of time, computed with a single direct integration.}
\end{figure}

\begin{figure}
\centering
\includegraphics[width=450pt]{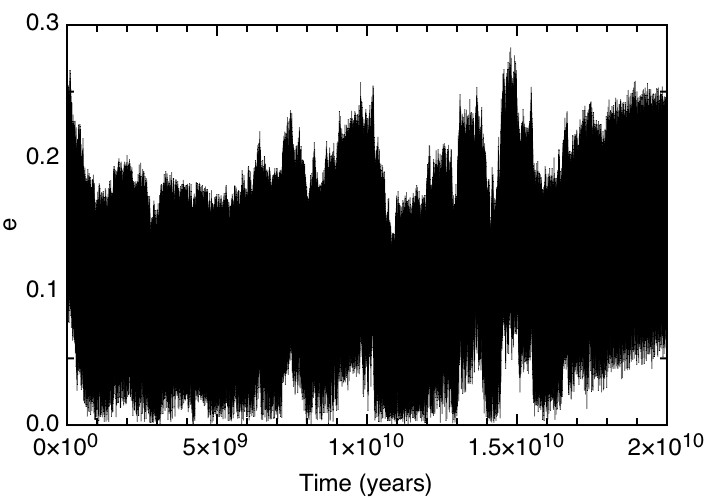}
\caption{Eccentricity of Mercury as a function of time, computed with a single direct integration.}
\end{figure}

\begin{figure}
\centering
\includegraphics[width=450pt]{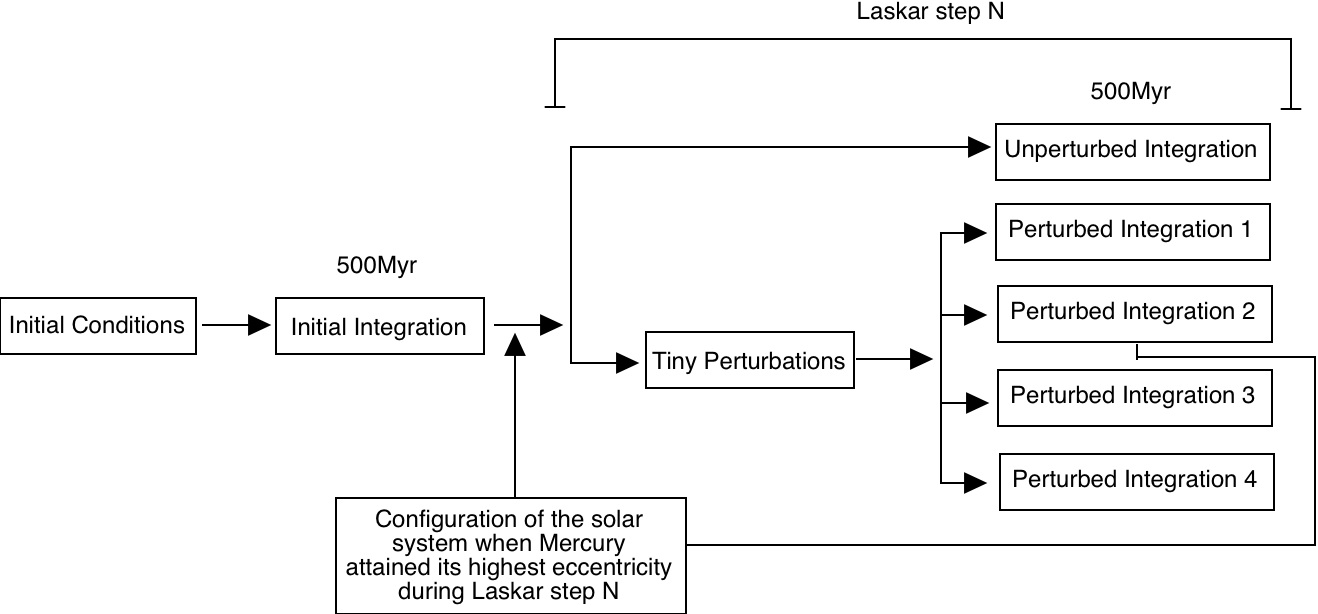}
\caption{A flow-chart summarizing our adaptation of the Laskar method.}
\end{figure}

\begin{figure}
\centering
\includegraphics[width=450pt]{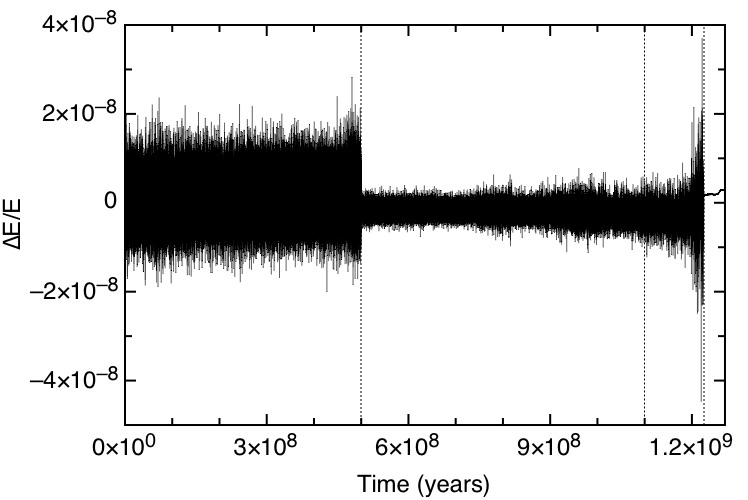}
\caption{Fractional change in total energy of the system, due to the integrator, as a function of time, in the 15m Laskar experiment. The dashed lines indicate where the integration time-step was reduced. Additionally, the third dashed line also indicates where we switched to the Bulirsch-Stoer algorithm from the symplectic algorithm. }
\end{figure}

\begin{figure}
\centering
\includegraphics[width=450pt]{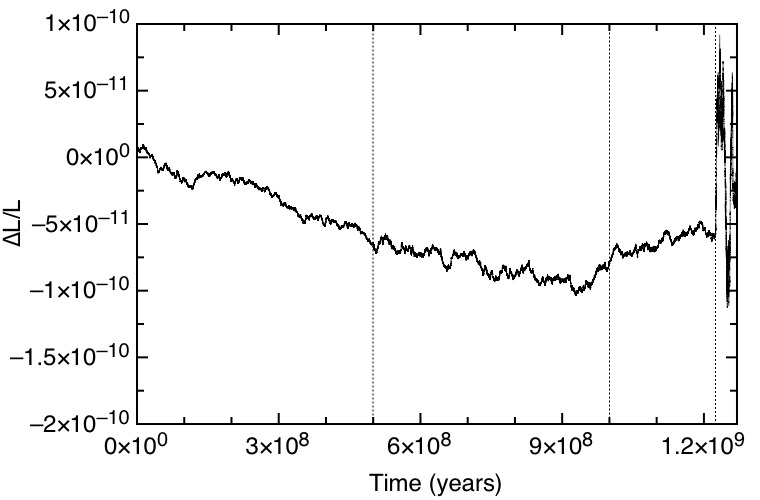}
\caption{Fractional change in total angular momentum of the system, due to the integrator, as a function of time, in the 15m Laskar experiment. The dashed lines indicate where the integration time-step was reduced. Additionally, the third dashed line also indicates where we switched to the Bulirsch-Stoer algorithm from the symplectic algorithm.}
\end{figure}

\begin{figure}
\centering
\includegraphics[width=450pt]{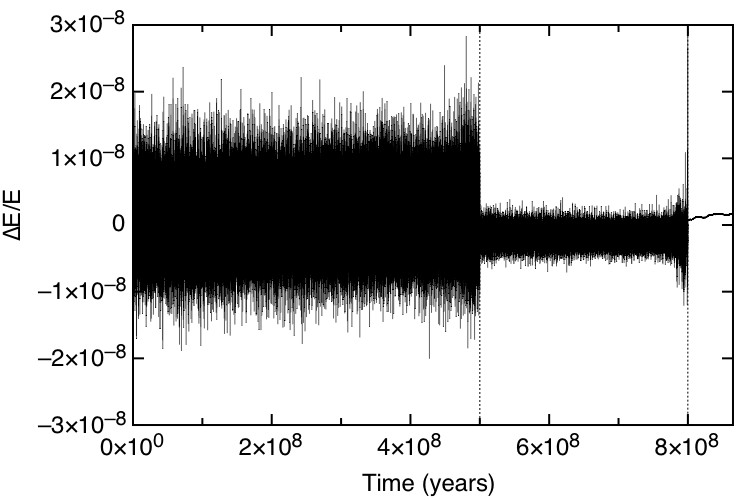}
\caption{Fractional change in total energy of the system, due to the integrator, as a function of time, in the 150m Laskar experiment. The dashed lines indicate where the integration time-step was reduced. Additionally, the second dashed line also indicates where we switched to the Bulirsch-Stoer algorithm from the symplectic algorithm. }
\end{figure}

\begin{figure}
\centering
\includegraphics[width=450pt]{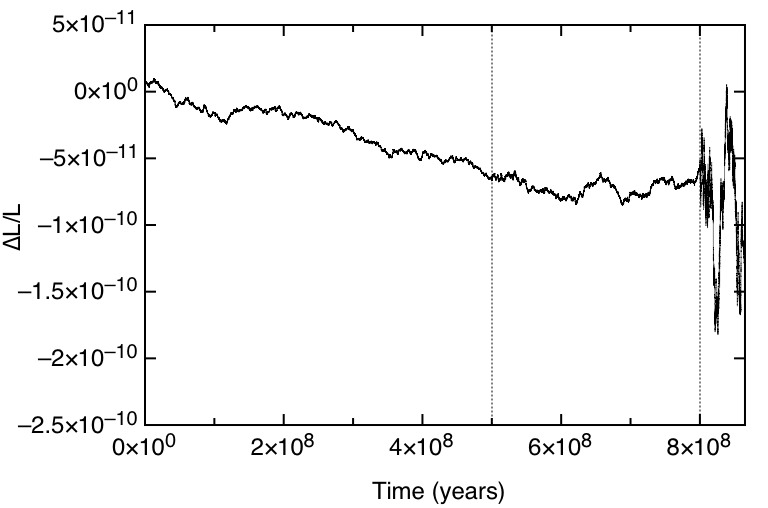}
\caption{Fractional change in total angular momentum of the system, due to the integrator, as a function of time, in the 150m Laskar experiment. The dashed lines indicate where the integration time-step was reduced. Additionally, the second dashed line also indicates where we switched to the Bulirsch-Stoer algorithm from the symplectic algorithm.}
\end{figure}

\begin{figure}
\centering
\includegraphics[width=450pt]{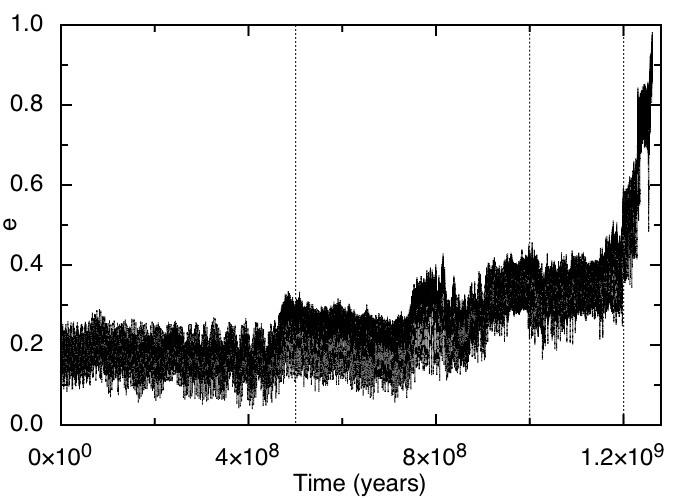}
\caption{Evolution of Mercury's eccentricity as a function of time in the 15m Laskar experiment. The dashed lines indicate the times at which the Laskar method was applied.}
\end{figure}

\begin{figure}
\centering
\includegraphics[width=450pt]{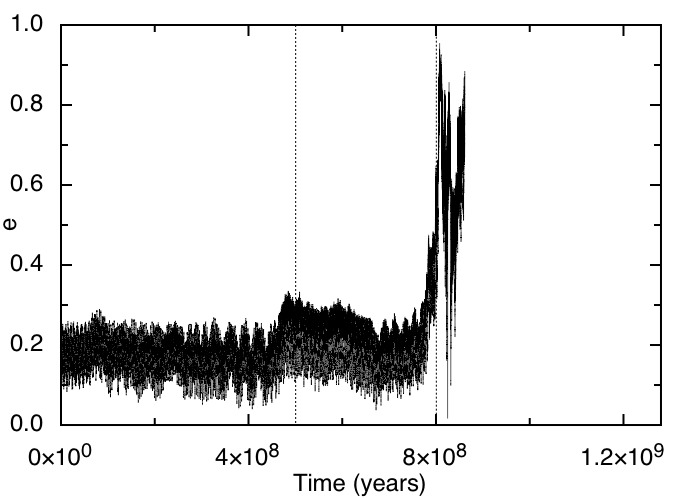}
\caption{Evolution of Mercury's eccentricity as a function of time in the 150m Laskar experiment. The dashed lines indicate the times at which the Laskar method was applied.}
\end{figure}

\begin{figure}
\centering
\includegraphics[width=450pt]{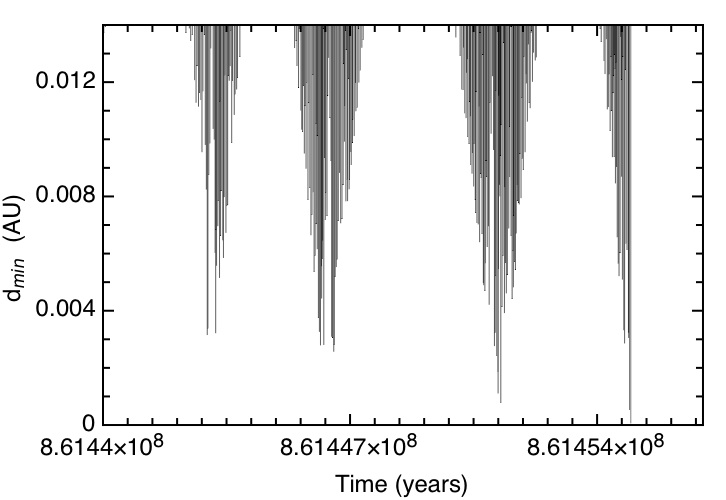}
\caption{The minimal distance of approach during a series of close encounters between Mercury and Venus as a function of time. The collision takes place at $t \sim$ 861.455Myr, when 
$d_{min} = 5.5561 \times 10^{-5}$ AU $< r_{venus} + r_{mercury} = 5.6762 \times 10^{-5}$ AU.}
\end{figure}

\begin{figure}
\centering
\includegraphics[width=450pt]{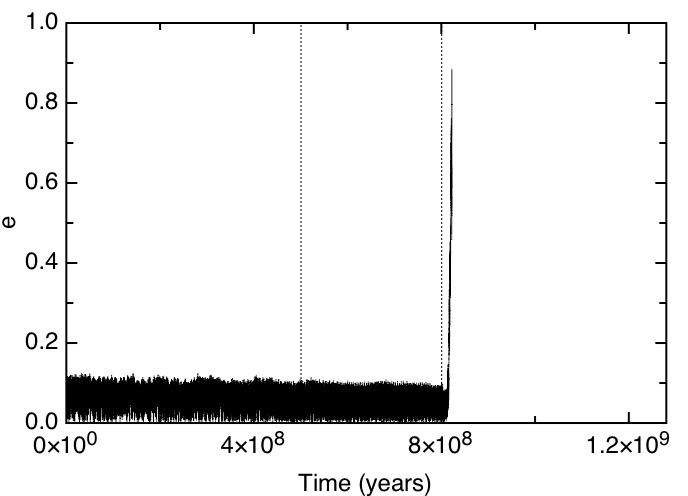}
\caption{Evolution of Mars' eccentricity as a function of time in the 150m Laskar experiment. The dashed lines indicate the times at which the Laskar method was applied. Note that in this solution, Mars escapes from the Solar System prior to Mercury's collision with Venus. Mars' escape is triggered by Mercury's entrance into a zone of greater chaos.}
\end{figure}

\begin{figure}
\centering
\includegraphics[width=450pt]{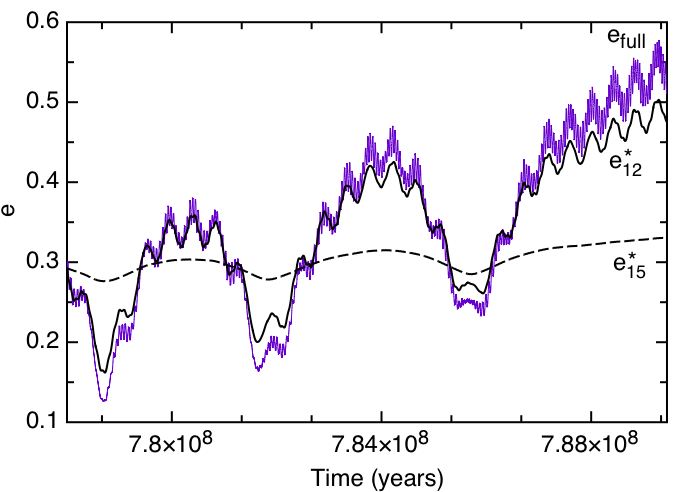}
\caption{Mercury's eccentricity as a function of time during the transition from stable to unstable motion in the 150m Laskar experiment. The blue curve represents actual Mercury's eccentricity obtained from numerical integration. The dashed curve of lower amplitude represents the partial solution $e^{*}_{15}$ obtained solely from the Mercury-Jupiter secular term of the Mercurian disturbing function. The black curve represents $e^{*}_{12}$, a partial solution for Mercury's eccentricity, obtained solely from the Mercury-Venus secular term in Mercury's disturbing function. }
\end{figure}

\begin{figure}
\centering
\includegraphics[width=450pt]{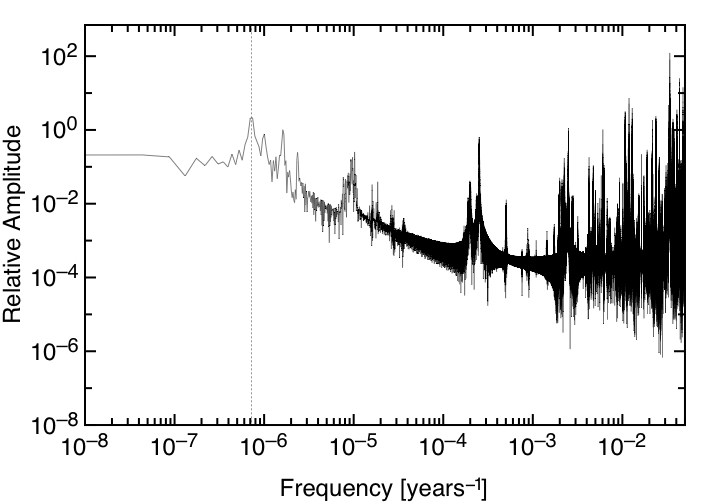}
\caption{A log-log plot of the Fourier analysis of Mercury's disturbing function corresponding to a stable configuration, in the frequency spectrum. The dashed line signifies the location of the frequency of the ($\varpi_{1}$ - $\varpi_{5}$) argument, where both $\varpi_{1}$ and $\varpi_{5}$ are measured numerically.}
\end{figure}

\begin{figure}
\centering
\includegraphics[width=450pt]{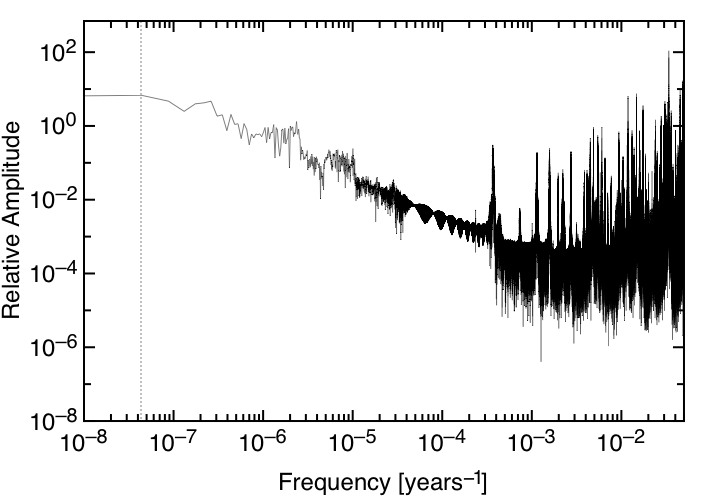}
\caption{A log-log plot of the Fourier analysis of Mercury's disturbing function during the transition from stable to unstable motion, in the frequency spectrum. The dashed line indicates the location of the frequency of the ($\varpi_{1}$ - $\varpi_{5}$) argument, where both $\varpi_{1}$ and $\varpi_{5}$) are measured numerically. Note that the ($\varpi_{1}$ - $\varpi_{5}$) apsidal angle has become resonant and shifted closer to zero compared with that of Figure (13). }
\end{figure}

\begin{figure}
\centering
\includegraphics[width=450pt]{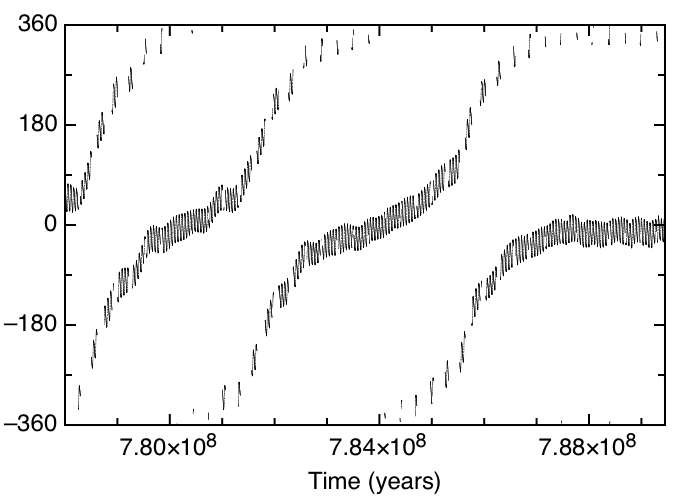}
\caption{The Mercury-Jupiter secular resonant argument ($\varpi_{1}$ - $\varpi_{5}$). The argument librates between +19.8 and -43.56 degrees from approximately 787.5Myr to 790.5Myr.}
\end{figure}

\begin{figure}
\centering
\includegraphics[width=450pt]{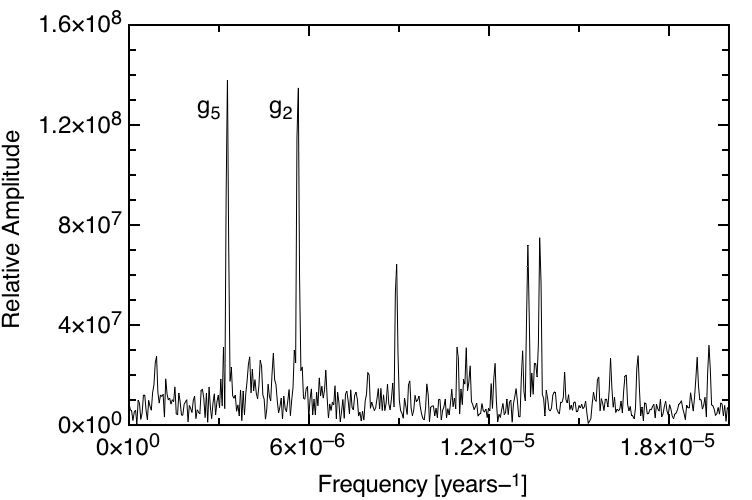}
\caption{The Fourier analysis of Venus's longitude of pericentre, $\varpi_{2}$. Note that aside from $g_{2}$, there exists a strong presence of the $g_{5}$ forcing mode.}
\end{figure}

\begin{figure}
\centering
\includegraphics[width=450pt]{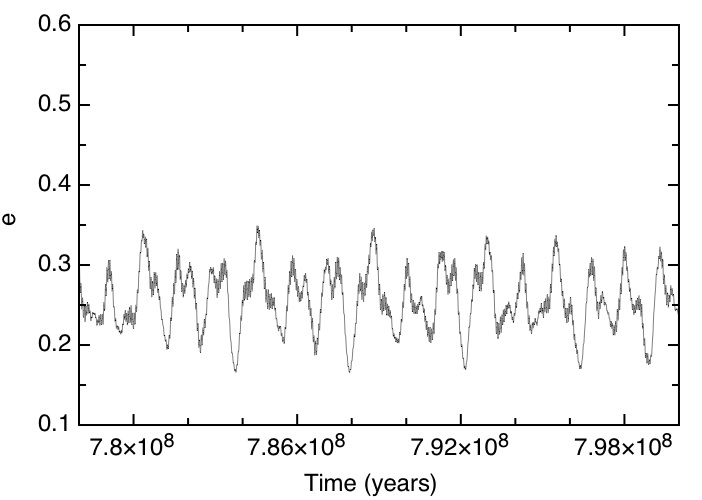}
\caption{Mercury's eccentricity as a function of time, taking into account the effects general relativity. Sudden introduction of general relativity stabilizes Mercury's behavior.}
\end{figure}

\begin{figure}
\centering
\includegraphics[width=450pt]{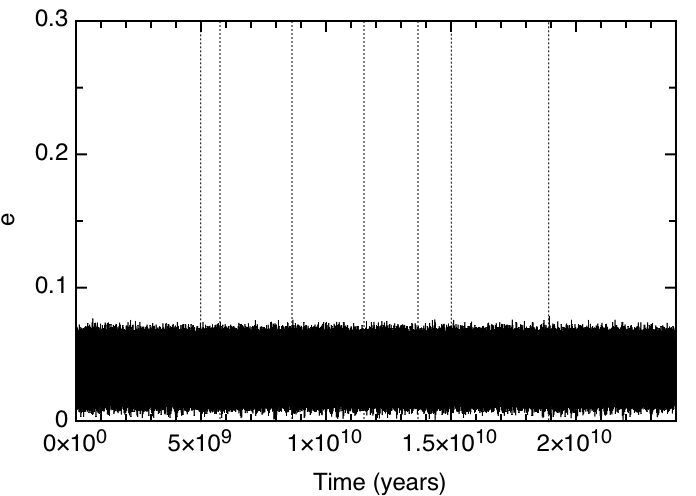}
\caption{Uranus' eccentricity as a function of time, in the Uranus Laskar experiment. The dashed lines indicate the times at which the Laskar method was applied.}
\end{figure}

\end{document}